\def\year{2022}\relax
\documentclass[letterpaper]{article} % DO NOT CHANGE THIS
\usepackage{aaai22}  % DO NOT CHANGE THIS
\usepackage{times}  % DO NOT CHANGE THIS
\usepackage{helvet}  % DO NOT CHANGE THIS
\usepackage{courier}  % DO NOT CHANGE THIS
\usepackage[hyphens]{url}  % DO NOT CHANGE THIS
\usepackage{graphicx} % DO NOT CHANGE THIS
\usepackage{natbib}  % DO NOT CHANGE THIS AND DO NOT ADD ANY OPTIONS TO IT
\usepackage{caption} % DO NOT CHANGE THIS AND DO NOT ADD ANY OPTIONS TO IT
\DeclareCaptionStyle{ruled}{labelfont=normalfont,labelsep=colon,strut=off} % DO NOT CHANGE THIS
\usepackage{algorithm}
\usepackage{algorithmic}

\usepackage{graphicx} %插入图片
\usepackage{booktabs}
\usepackage{multirow}

\usepackage{newfloat}
\usepackage{listings}
\floatstyle{ruled}
\newfloat{listing}{tb}{lst}{}
\floatname{listing}{Listing}
\title{FORCE: A Framework of Rule-Based Conversational Recommender System}
\author {
        Jun Quan\textsuperscript{\rm 1,2},
        Ze Wei\textsuperscript{\rm 1},
        Qiang Gan\textsuperscript{\rm 1} \thanks{Corresponding author is Qiang Gan(qigan@microsoft.com).}, 
        Jingqi Yao\textsuperscript{\rm 1,3}, 
        Jingyi Lu\textsuperscript{\rm 1}, 
        Yuchen Dong\textsuperscript{\rm 1}, 
        Yiming Liu\textsuperscript{\rm 1}, 
        Yi Zeng\textsuperscript{\rm 1}, 
        Chao Zhang\textsuperscript{\rm 1}, 
        Yongzhi Li\textsuperscript{\rm 1}, 
        Huang Hu\textsuperscript{\rm 1}, 
        Yingying He\textsuperscript{\rm 1}, 
        Yang Yang\textsuperscript{\rm 1} and Daxin Jiang\textsuperscript{\rm 1}\\
}

\affiliations {
    \textsuperscript{\rm 1} Microsoft, China \\
    \textsuperscript{\rm 2} School of Computer Science and Technology, Soochow University, Suzhou, China \\
    \textsuperscript{\rm 3} Donald Bren School of Information and Computer Sciences, University of California, Irvine\\
    $\{$junquan, zewei, qigan, jinglu, yucdong, yiml, yize, chaozh, yongli, huahu, yingyhe, yayan, djiang$\}$@microsoft.com,  jessieyao0602@gmail.com
}

\begin{document}

\maketitle

\begin{abstract}
The conversational recommender systems (CRSs) have received extensive attention in recent years. However, most of the existing works focus on various deep learning models, which are largely limited by the requirement of large-scale human-annotated datasets. Such methods are not able to deal with the cold-start scenarios in industrial products. To alleviate the problem, we propose {\bf FORCE}, a {\bf F}ramework {\bf O}f {\bf R}ule-based {\bf C}onversational r{\bf E}commender system that helps developers to quickly build CRS bots by simple configuration. We conduct experiments on two datasets in different languages and domains to verify its effectiveness and usability. 
\end{abstract}

%\vspace{-0.2cm}
\section{Introduction}
\label{Introduction}
Recommender system is a topic undergoing intense academic research and industrial production. The traditional recommender systems usually work statically and cannot obtain the latest explicit user feedback. So the conversational recommender systems (CRSs) come into being \cite{gao2021advances}. CRSs obtain users' real-time feedback through multi-turn dialogue and make recommendations \cite{zhou-etal-2021-crslab}. However, existing methods usually need massive annotated data for each scenario and mainly focus on movie and music \cite{chen2019towards, zhou2020improving} due to the limitation of datasets \cite{li2018conversational, kang2019recommendation, hayati2020inspired, moon2019opendialkg, zhou2020towards, liu2020towards}. In fact, CRSs can be applied in many areas far more than these domains and have great potential in industrial products. 

Unlike existing highly customized models, we focus on a different track to build a configurable framework that can support developers in constructing low-cost cold-start CRS bots. We introduce {\bf FORCE}, a {\bf F}ramework {\bf O}f {\bf R}ule-based {\bf C}onversational r{\bf E}commender system.

We propose to implement a modularized CRS by reasoning on a simplified knowledge graph (KG) and mapping the bot decisions to three types of dialogue acts, in the form of tree structure beginning with three bot intents (\emph{Query / Recommend / Chat}), inspired by the CR-Walker \cite{ma2020bridging}. The framework can quickly build CRS bots and require only a data file for KG construction and some configurations for style. 
To increase the usability of our framework, we provide two pre-built modes: \emph{casual} and \emph{cautious}. \emph{Casual} mode is a broad recommendation mode, suitable for the recommendation in entertainment scenarios such as movies, music, and books. \emph{Cautious} mode is a more precise recommendation mode for medical diagnosis, psychological counseling, financial consultation, etc. The style and policy of the bot can be further adjusted in each mode.
Besides, the FORCE is language-independent and can easily adapt to various linguistic markets.

% To verify the effectiveness and usability, we build CRS bots on an English movie-domain dataset and a Chinese diagnosis-domain dataset respectively. Both bots produce promising evaluation results.

The contributions of our work are as follows. 1) We propose FORCE, a framework of rule-based conversational recommender system, which facilitates the construction of CRS bots with simple configurations. 2) We verify the effectiveness of FORCE with two scenarios in different languages, and both CRS bots produce promising evaluation results. 3) We provide a demonstration video describing the system architecture and modularization, which will be valuable for developers in building real-world CRS applications \footnote{Demo video: \url{https://youtu.be/UjkLL5pgXx8}}.

\begin{figure}[tp!] %图片浮动环境，类似表格中的 table [htp] 参数和表格的类似
\centering %图片居中
\includegraphics[scale=0.28]{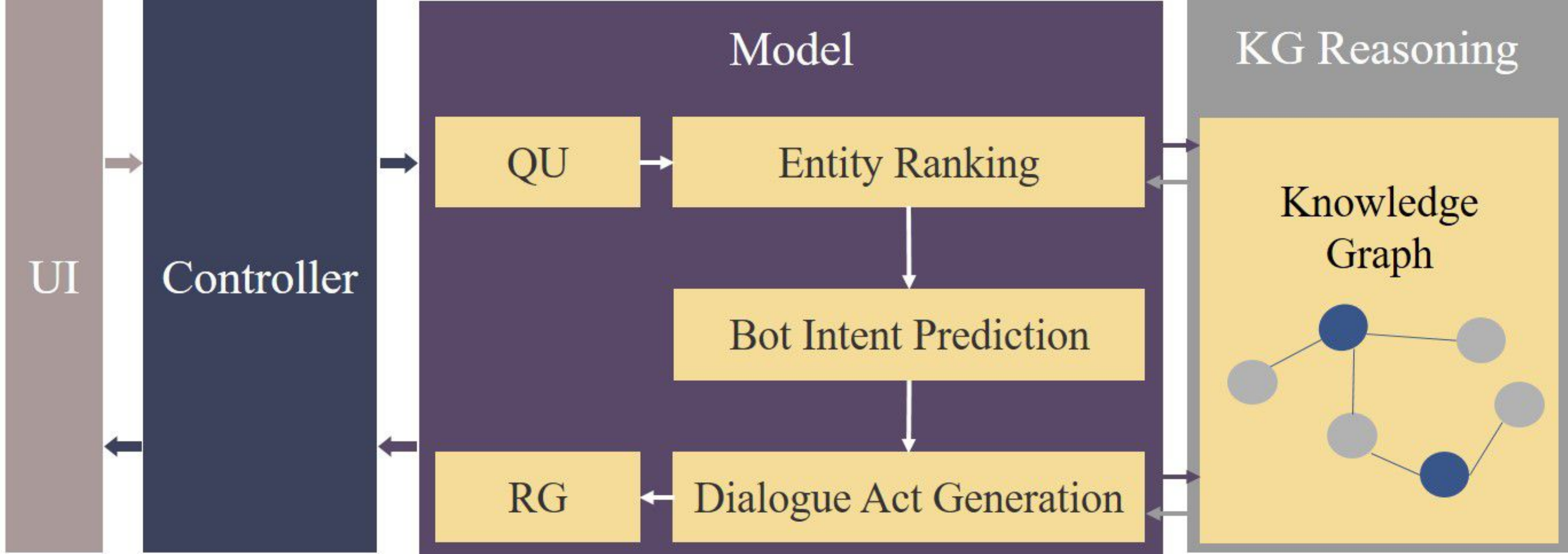}%图片大小，文件名
\caption{The overall architecture of the FORCE.}%图片的标题
\label{architecture} %引用
\end{figure}

%\vspace{-0.2cm}
\section{System Overview}

The overall architecture and workflow of FORCE are shown in Figure \ref{architecture}. The system is designed based on the Model-View-Controller pattern and consists of 4 different layers.

The User Interface Layer provides a user-friendly interactive environment, making it convenient for developers to build a CRS bot and have a multi-turn conversation with it.
The Controller Layer is responsible for overall scheduling, receiving, and forwarding messages between the front-end user interface and the back-end model.

The Model Layer implements a modularized workflow, including Query Understanding (QU), Entity Ranking, Bot Intent Prediction, Dialogue Act Generation, and Response Generation (RG). (1) The QU module parses user query into a \emph{semantic frame}, a structured representation of user intent and slots, based on configuration. (2) The Entity Ranking module ranks all candidate entities according to conversational context. Entity neighborhood and mentioned attributes on KG are both considered in the ranking. (3) The Bot Intent Prediction module decides which action the bot should take at the current turn based on \emph{semantic frame} and conversational context. ``Condition-Decision'' rules are implemented in this module. (4) Based on the outputs of previous modules, the Dialogue Act Generation module further infers on the knowledge graph to generate a structured representation of the next action the system should take, called the \emph{Dialogue Act}. (5) The Response Generation (RG) module converts the \emph{Dialogue Act} into the bot response in natural language form based on configured templates \footnote{Response template example with placeholder: ``Which director's movies do you like ? E.g. \{attributes\}''}.

The Knowledge Graph Reasoning Layer is responsible for connecting the system to a simplified knowledge graph, which is constructed by the uploaded data file. It provides various reasoning algorithms on KG, and serves Entity Ranking and Dialogue Act Generation module.

It's worth mentioning that the modules are decoupled by pre-defined interfaces, thus each module can be replaced with a machine learning model or service. We leave the support of uploading annotated data for model training and service configuration on framework to future work.

\section{Demonstration}
\label{Demonstration}
The process of building a CRS bot can be divided into the following three stages, as shown in our demo video.

\noindent \textbf{KG Configuration:} Developers need to choose an existing KG or upload a new data file for KG construction. Each line of data file represents one connection between an attribute and an entity or a generic node \footnote{Data example: ``Catch Me If You Can[tab]Entity[tab]Tom Hanks[tab]Attribute[tab]Actor''}. 
% Figure \ref{kg} is a sample tsv file for movie KG.

\noindent \textbf{Bot Configuration:} Developers then configure the style of the bot. The preferences of `\emph{Query / Recommend / Chat}' can be changed through the scroll bars. The effect to users' denial can be adjusted by \emph{negation penalty}. \emph{Matching threshold} controls the recommendation caution degree. Additionally, the keywords for Query Understanding and templates for Response Generation are configurable.

\noindent \textbf{Bot Interaction:} After the two stages above, a CRS bot is built. The portal includes the chat window and bot status. The \emph{semantic frame} parsed by QU module is displayed under each user utterance. Based on conversation context, the bot reasons the KG and decides the next action. 
The real-time visualization of the KG is designed for interpretability.

Our demo system clearly demonstrates the workflow of the decision-making and reasoning process, making the CRS more transparent, interpretable, and controllable. Those information can help developers to further improve the system.

% \begin{table}[t!]
% \setlength{\abovecaptionskip}{0.2cm}
% \setlength{\belowcaptionskip}{-0.3cm}
% \centering
% \scalebox{0.55}{
% \begin{tabular}{@{}lccccccc@{}}
% \toprule
% \multicolumn{1}{c}{\multirow{2}{*}{Dataset}}& \multirow{2}{*}{Domain} & \multirow{2}{*}{Language}  & \multirow{2}{*}{Dialogues} & \multirow{2}{*}{Turns} & \multicolumn{3}{c}{Knowledge Graph} \\ \cmidrule(l){6-8} 
% \multicolumn{1}{c}{}                        &                             &                             &                            &                        & Nodes & Edges & Entities \\ \midrule
% M-RD                                        & Movie                       & English                     & 54                         & 324                    & 2667  & 8908  & 246      \\
% DX                                          & Medical                     & Chinese                     & 527                        & 1408                   & 47    & 109   & 5        \\ \bottomrule
% \end{tabular}
% }
% \caption{Dataset Statistics.}
% \label{dataset}
% \end{table}

\begin{table}[t!]
%\normalsize
\large
\centering
\scalebox{0.66}{
\begin{tabular}{llcccc}
\toprule
\multirow{2}{*}{Dataset} & \multirow{2}{*}{Method} & Intent Prediction & \multicolumn{3}{c}{Entity Recommendation} \\ \cline{3-6} 
                         &                         & Accuracy (\%)         & R@1 (\%)    & R@10 (\%)    & R@50 (\%)    \\ \hline
\multirow{2}{*}{M-RD}    & Baseline      & 33.33                 & 0.41        & 4.10         & 20.30        \\
                         & Ours                    & 67.28                 & 4.37        & 16.50        & 50.97        \\ \hline
                         &                         & Accuracy (\%)         & R@1 (\%)    & R@2 (\%)     & R@3 (\%)     \\ \hline
\multirow{2}{*}{DX}      & Baseline      & 33.33                 & 20.00       & 40.00        & 60.00        \\
                         & Ours                    & 66.34                 & 79.51       & 92.98        & 97.15        \\ \bottomrule
\end{tabular}
}
\caption{Experimental results on M-RD and DX datasets.}
\label{results}
\end{table}

%\vspace{-0.2cm}
\section{System Evaluation}

%We conduct experiments on two datasets named M-RD and DX which are sampled and reprocessed from open datasets %\cite{li2018conversational,ma2020bridging,xu2019end} and cover different domains. Simplified KGs are constructed for them.

We conduct experiments in two different domains and languages to evaluate the usability. For movie domain, we sample the conversations \footnote{We sample conversations containing 5-7 turns with all \emph{Query / Recommend / Chat} intents.} from ReDial \cite{li2018conversational,ma2020bridging} and form M-RD (Mini-Redial). For medical domain, we reprocess a Chinese medical diagnosis dataset DX  \cite{xu2019end}. Data files to construct the simplified KG are prepared based on the datasets.

We evaluate the \emph{Accuracy} for bot intent prediction task at each turn and the \emph{Recall@k} for recommendation task at the turns when the bot intent is `\emph{Recommend}'.
With the implementation and configuration shown in our demo video, we get the experiment results in Table \ref{results}. We take the theoretical results of random sampling as baseline, in order to have a fair comparison with cold-start situations. 
For the bot on movie domain, the results are promising compared with baseline. CR-Walker \cite{ma2020bridging} can also be used as a reference point \footnote{CR-Walker on full Redial gets Accuracy: 67.8\%, Entity Recall: 3.1\%(R@1) / 15.5\%(R@10) / 36.5\%(R@50).}
. For the bot about medical diagnosis, we expect the bot to make the recommendation as accurately as possible. The results of \emph{Recall@k} show that our bot can provide reliable suggestions \footnote{We add medical diagnosis to verify the domain scalability besides common entertaining scenarios. Ethical issues should be considered when building real-world medical applications.}.

We share a similar conclusion with \citet{ma2020bridging} that a different bot intent from the human annotator does not indicate a poor choice. For bots built by our framework, it's more about the configured styles by developers.

\section{Conclusion}

In this paper, we introduce {\bf FORCE}, a framework of rule-based conversational recommender system. It supports developers in constructing low-cost cold-start CRS bots. The framework is language and domain independent. We hope it can not only help build early-stage CRS products, but also serve as a conversation collection tool for data-driven improvements later. 
% We will make continuous efforts to improve it in future. On one hand, we will provide more built-in configurations in different languages for development convenience. On the other hand, we will display more details on the decision making and graph reasoning process for interpretability and debugging.

% \section{Acknowledgments}
% This work is supported by Microsoft STCA NLP Group. We would also like to thank the anonymous reviewers for their insightful comments.

\bibliographystyle{IEEEtran}
% \bibliography{main}

\end{document}